\documentclass[a4paper,11pt]{article}

\usepackage[utf8]{inputenc}
\usepackage[T1,T2A]{fontenc}
\usepackage{amsmath,amsfonts,amssymb}
\usepackage{cite}

\usepackage{geometry}
\numberwithin{equation}{section}

\newcommand{\lb}{\left(}
\newcommand{\rb}{\right)}
\newcommand{\lsb}{\left[}
\newcommand{\rsb}{\right]}
\newcommand{\veps}{\varepsilon}
\newcommand{\pz}{\partial_z}
\newcommand{\pzz}{\partial^2_{zz}}
\newcommand{\hq}{\hat{Q}}

\allowdisplaybreaks

\begin{document}

\renewcommand*{\thefootnote}{\fnsymbol{footnote}}

\begin{center}
{\large\bf On incorporation of heavy-quark mass into soft-wall holographic models}
\end{center}
\begin{center}
{S.S. Afonin\(^{a,b}\)\footnote{E-mail: \texttt{s.afonin@spbu.ru}} and
T.D. Solomko\(^{c}\)
}
\end{center}

\renewcommand*{\thefootnote}{\arabic{footnote}}
\setcounter{footnote}{0}

\begin{center}
  {\small\({}^a\)Saint Petersburg State University, 7/9 Universitetskaya nab.,
  St.Petersburg, 199034, Russia}\\
  \vspace*{0.15cm}
  {\small\({}^b\)National Research Center "Kurchatov Institute": Petersburg Nuclear Physics Institute,
  mkr. Orlova roshcha 1, Gatchina, 188300, Russia}\\
  \vspace*{0.15cm}
  {\small\({}^c\)Department of Physics, Ben-Gurion University of the Negev, Beer-Sheva 84105, Israel}
\end{center}


\begin{abstract}

We consider the soft-wall holographic model with the linear dilaton background.
The model leads to a Hydrogen-like meson spectrum which can be interpreted as the static limit with very large quark masses 
when the Coulomb interaction dominates. The mass scale introduced by the linear dilaton is
matched to the quark mass. The resulting model is analyzed for the scalar, vector and tensor cases.
The electromagnetic coupling constants predicted by the model are decreasing with the radial
number in contrast to the soft-wall model with quadratic dilaton where these couplings represent a universal constant.
The given prediction is qualitatively consistent with the corresponding experimental data in vector quarkonia.
The proposed model can thus be used as a constituent part of more elaborated holographic models for heavy quarkonia.
A particular example of such a model is put forward.

\end{abstract}

\bigskip

\section{Introduction}

The holographic hypothesis of anti-de Sitter/conformal field theory (AdS/CFT) correspondence~\cite{mald,witten,gub} has become a cornucopia of countless
inspirations in theoretical physics over the past 25 years. Being a purely string-theoretic construction,
the idea of this correspondence led to the appearance of numerous phenomenological approaches in which strongly coupled
quantum systems are described in terms of fields propagating in extradimensional curved spaces.
A big part of those phenomenological applications was developed for modeling the non-perturbative strong interactions (the so-called AdS/QCD approach).
There are two different kinds of AdS/QCD models: top-down (based on some brane constructions in string theories which lead to low-energy gauge
theories somewhat similar to QCD) and much more numerous bottom-up models (based on incorporation of basic properties of QCD into a five-dimensional
phenomenological frameworks by specifying the geometry of modified or truncated AdS space and some bulk fields).

Within the bottom-up holographic approach, the most popular type of models are variations of the Soft-Wall (SW) holographic model
proposed in~\cite{son2,andreev}. The 5D action of these models is usually written in the static AdS$_5$ space and includes a $z$-dependent dilaton
background ($z$ is the holographic coordinate) that introduces a mass scale and dictates the physics in the infrared.
This phenomenological approach to strong interactions has various heuristic motivations~\cite{Afonin:2022qby} and
by now it counts several hundreds of publications on variety of applications in hadron physics
(Regge spectroscopy, hadron formfactors, QCD phase diagram, etc.~\cite{Afonin:2021cwo}).
As a rule, the constructed holographic models describe the sector of light quark flavors, while
the progress in holographic description of heavy flavors looks relatively modest. For instance, the SW description of the
experimental spectra of heavy quarkonia was addressed only in a few papers~\cite{Kim:2007rt,Grigoryan:2010pj,Branz:2010ub,Gutsche:2012ez,Afonin:2013npa,Braga:2015lck,Rinaldi:2020ssz,MartinContreras:2021bis,Chen:2021wzj}.
Perhaps an underlying reason is that the spectrum of excited light mesons has a clear expected pattern in the first approximation ---
a linear Regge form (see, e.g., discussions in~\cite{ani,bugg,phen4,phen5,klempt,phenSV,arriola,Afonin:2020bqc,Chen:2021kfw} and references therein).
One can easily identify a 5D background
that reproduces this form of spectrum, which is the quadratic exponential background of Refs.~\cite{son2,andreev}, and then use this background
(or its modifications) for developing a hadron phenomenology of interest. In the heavy quarkonia, however, the first approximation
to the form of observed spectrum is not obvious. The problem is exacerbated by the lack of a well-founded holographic recipe
for incorporating the quark masses into the bottom-up AdS/QCD models. For the light flavors, the given trouble is avoided by
setting the quark masses to zero. In the aforementioned papers~\cite{Kim:2007rt,Grigoryan:2010pj,Branz:2010ub,Gutsche:2012ez,Afonin:2013npa,Braga:2015lck,Rinaldi:2020ssz,MartinContreras:2021bis,Chen:2021wzj},
the masses of heavy quarks were inserted into some SW-like models
using various {\it ad hoc} phenomenological prescriptions. The purpose of the present work is to motivate a universal prescription
for incorporating the quark mass scale into the SW holographic models.

We construct a SW-like holographic model in the limit of very large quark mass where the dynamical QCD scale
(usually related with the gluon condensate and incorporated into SW models via the quadratic dilaton background) can be neglected.
In the non-relativistic approximation, the Coulomb part of confining potential in heavy quarkonia is known to become important.
In the limit when the Coulomb interaction absolutely dominates, the spectrum should have a Hydrogen-like form.
This type of spectrum is reproduced if the quadratic dilaton background is replaced by the linear background.
We will show that the mass scale introduced by the linear dilaton can be naturally related to the quark mass.

Certainly the SW model with linear dilaton (called a ``linear SW model'' in what follows) cannot describe the real spectrum
of heavy quarkonia but we believe that the action of this model
should be considered as an important component in the construction of realistic SW holographic models for heavy flavors.
Aside from this aspect, the proposed model may be interesting by itself. The discrete spectrum of the standard SW model with
quadratic dilaton is determined by a one-dimensional Schr\"{o}dinger equation with a potential of harmonic oscillator.
The resulting spectrum has a certain form of degeneracy between the radial and orbital excitations. It is well known that
only two quantum-mechanical problems reveal this kind of degeneracy: the harmonic oscillator and the Coulomb potential~\cite{landau}.
A question appears what kind of bottom-up holographic model corresponds to the second possibility and how these two possibilities are
related? The linear SW model provides the answer. We note that the eigenfunctions in both problems are expressed
via the associated Laguerre polynomials, this may indicate that the description of hadron structure in the framework of both
models could be not very different. In addition, as we demonstrated in~\cite{universe}, the Cornell-like confinement potential obtained in the
linear SW model from the holographic Wilson loop is quantitatively close to the one obtained in the quadratic SW model.

The relevance of the linear SW model to the light flavors is also not excluded. The experimental spectrum of light hadrons has
saturation scale, $M_\text{sat}\approx2.5$~GeV~\cite{Afonin:2006vi,Szanyi:2023ano}, above which the light resonances are practically not observed. A possible interpretation
of this phenomenon is the hadron string breaking. In one of realizations of this idea, the spectrum becomes purely coulombian at
sufficiently large quark-antiquark distances~\cite{gonzales}. One can imagine that experimentally the Hydrogen-like condensation of energy levels
can hardly be detected because the splitting between masses of hadron excitations becomes much smaller than their decay width, so resonances
just merge into continuum immediately after the coulombian regime sets in near $M_\text{sat}$. Thus the linear SW model proposes
a specific way for incorporation of the saturation scale $M_\text{sat}$ into the bottom-up holographic framework, possibly as a part
of more general AdS/QCD models.

From the viewpoint of AdS/CFT correspondence in string theory, the linear SW holographic model may be of interest
as a phenomenological toy model in the following sense.
Quite recently Maldacena and Remmen analyzed the amplitude of the scattering of open strings attached to a D-brane in the AdS space~\cite{Maldacena:2022ckr}.
It turned out that its infinite spectrum of massive states exhibits an accumulation point below a certain energy
resembling that in the Hydrogen atom\footnote{The accumulation points might be a generic quality of consistent gravitational effective
field theories, i.e., the theories subject to constraints implied from unitarity, causality and crossing-symmetry.
The recent analysis of the space of such theories performed in Ref.~\cite{Chiang:2022jep} showed that
the majority of the region is populated by spectrums with an accumulation point.}.
Moreover, the radial spectrum reproduced closely the spectrum of the long-known Coon dual amplitude~\cite{coon,coon2} --- a unique one-parameter extension of
the famous Veneziano amplitude~\cite{veneziano} that preserves its analytical properties and leads to a unitary $S$-matrix\footnote{Speaking more
precisely, the Coon dual amplitude represents the $q$-deformation of the Veneziano amplitude, it smoothly interpolates between the
Veneziano amplitude as $q\rightarrow1$ and scalar field theory as $q\rightarrow0$, recent discussions are given in~\cite{volovich,Jepsen:2023sia}.
The Coon amplitude generalizes the structure of Veneziano amplitude as infinite product of poles with crossing symmetry but leads to logarithmic Regge trajectories.
Note that the appearance of an infinite spectrum of massive states approaching a finite accumulation point have also been found in
several recent amplitude studies, see the discussions and references in~\cite{Maldacena:2022ckr,volovich,Jepsen:2023sia}.}.
The quadratic SW holographic model reproduces the spectrum of Veneziano amplitude~\cite{son2} that is typical for scattering of
strings in a flat space. In this sense, the standard quadratic SW model may be viewed as a toy model for this situation on the hadron level.
In the same sense, the linear SW model might be interpreted as a rough toy model for the spectrum of open string excited in the AdS space.
The word ``rough'' in this context means that an exponential approaching to the accumulation point found in~\cite{Maldacena:2022ckr} (and taking place
in the Coon amplitude) is replaced by a polynomial approaching in our holographic model.

The paper is organized as follows. In Section~2, we remind the reader the structure of the simplest quadratic SW holographic model
and introduce the linear SW holographic model in the scalar case. In Section~3, the model is extended to the vector case and the correlator
of vector currents is calculated. The model is further generalized to arbitrary integer spins in Section~4, some phenomenological modifications
are also discussed. The incorporation of heavy-quark mass is analyzed in Section~5. An example of more realistic extension of the model describing 
the heavy quarkonia is briefly considered in Section~6. Our conclusions are presented in the final Section~7. 
The basic reference formulas on the Coulomb problem used in the text are given in the Appendix.

\section{Linear SW model in the scalar case}

The action of the standard SW model is given by (a general normalization factor is omitted)~\cite{son2}
\begin{equation}
\label{sw}
S=\int d^5x \sqrt{g} \, e^{f}\mathcal{L},
\end{equation}
where a specific SW model is defined by a specific Lagrangian $\mathcal{L}$ and by the function $f(z)$ in the exponential
background. By analogy with string theory, the latter is often called the ``dilaton background''.
Following the AdS/CFT correspondence~\cite{mald,witten,gub}, the model is formulated in the 5D anti-de Sitter (AdS$_5$) space,
more exactly, in the Poincar\'{e} patch of AdS$_5$ space. The non-zero components of the corresponding
metric are
\begin{equation}
\label{metric}
  g_{\mu\nu}=\frac{R^2}{z^2}\,\eta_{\mu\nu},\qquad
  g_{zz}=-\frac{R^2}{z^2},\qquad
  \eta_{\mu\nu}=\text{diag}\left\lbrace 1,-1,-1,-1\right\rbrace,
\end{equation}
therefore,
\begin{equation}
\label{det}
\sqrt{g}=\sqrt{|\text{det}g_{MN}|}=\frac{R^5}{z^5}.
\end{equation}
Here $R$ is the radius of (AdS$_5$) space.
In the original SW model of Ref.~\cite{son2}, the function $f(z)$ is quadratic in the holographic coordinate $z$,
\begin{equation}
\label{SWbac}
f(z)=-\lambda^2 z^2,
\end{equation}
with the constant $\lambda$ introducing the mass scale. This simple SW model leads to the linear Regge spectrum,
\begin{equation}
\label{spSW}
 M^2_{n,J}=4\lambda^2(n+J), \qquad n=0,1,\dots, \qquad J=1,2,\dots.
\end{equation}
For zero spin, $J=0$, the relation~\eqref{spSW} is modified by a constant contribution, $M^2_{n,0}=4\lambda^2(n+\text{const})$.
The discrete spectrum~\eqref{spSW} arises when the 5D fields in $\mathcal{L}$ are dual on the 4D boundary of AdS$_5$
to twist-2 (i.e., the mass dimension minus spin) operators in QCD (except the scalar case, where the twist must be
higher and this results in the aforementioned constant contribution). An extension to higher twists was considered
in~\cite{afonin2020}, it just adds some constants to the spectrum~\eqref{spSW}.

It is noteworthy that the holographic SW spectrum~\eqref{spSW} reproduces the spectrum of Veneziano dual
amplitude~\cite{veneziano} which gave rise to the modern string theory.

Let us consider in~\eqref{sw} the linear background function
\begin{equation}
\label{SWbac2}
f(z)=2cz,
\end{equation}
the factor of \(2\) here is inserted to simplify our further formulas. The confining properties of SW model with this
background (which take place at positive mass parameter, $c>0$) were discussed in~\cite{universe} and the corresponding Cornell-like
potential was calculated. In the simplest case, the Lagrangian $\mathcal{L}$ contains just a free real scalar field,
the linear SW model takes the form
\begin{equation}
\label{act}
  S=\int d^5x \sqrt{g}\, e^{2cz}\left(g^{MN}\partial_M\Phi\partial_N\Phi-m_5^2\Phi^2\right).
\end{equation}
According to the AdS/CFT dictionary~\cite{witten,gub}, the 5D mass $m_5$ of a scalar field in AdS$_5$
is related with the canonical dimension $\Delta$ of corresponding scalar operator in dual
4D field theory as
\begin{equation}
\label{s_delta}
m_5^2R^2=\Delta(\Delta-4).
\end{equation}
For instance, if $\Phi$ is dual to the QCD operator $\alpha_s G_{\mu\nu}^2$, then $m_5=0$.

The equation of motion following from~\eqref{act},
\begin{equation}
  \partial_N\lb e^{2cz}\sqrt{g}\,g^{MN}\partial_M\Phi\rb+e^{2cz}\sqrt{g}\,m_5^2\Phi=0,
\end{equation}
in the case of a curved space with metric~\eqref{metric}, becomes
\begin{equation}
\label{s_eom0}
  \partial^\mu\partial_\mu\Phi-z^3e^{-2cz}\pz\lb\frac{e^{2cz}}{z^3}\pz\Phi\rb+
  \frac{R^2}{z^2}m_5^2\Phi=0.
\end{equation}
The substitution \(\Phi=z^{3/2}e^{-cz}\phi\) transforms~\eqref{s_eom0} into a simpler form,
\begin{equation}
\label{s_eom}
  \partial^\mu\partial_\mu\phi-\pzz\phi+
  \lb\frac{15/4+m_5^2R^2}{z^2}-\frac{3c}{z}+c^2\rb\phi=0.
\end{equation}
The free particles in flat 4D space are described by the plane waves, this dictates
the standard holographic ansatz for a particle-like excitation of mass $M$,
\begin{equation}
\label{pe}
  \phi(x,z)=e^{iqx}\psi(z),\qquad q^2=M^2.
\end{equation}
This ansatz in~\eqref{s_eom} results in a one-dimensional Schr\"{o}dinger equation,
\begin{equation}
\label{eom}
  -\psi''+V\psi=q^2\psi,
\end{equation}
with the potential
\begin{equation}
\label{eom_pot}
  V(z)=\frac{15/4+m_5^2R^2}{z^2}-\frac{3c}{z}+c^2.
\end{equation}
One can immediately recognize the classical Schr\"{o}dinger equation for the
radial part of wave function in the Coulomb problem\footnote{To avoid a confusion, we recall that the potential~\eqref{eom_pot} is
obtained in the supposed limit of infinitely heavy quark mass. A realistic model for heavy quarkonia should contain growing with $z$
part(s) responsible for the confining effect. A particular model is briefly considered in Section~6.}.
The corresponding solution is briefly recapitulated in the Appendix.
Making use of the formulas from Appendix, one can immediately write out the discrete spectrum,
\begin{equation}
  M_n^2=c^2-\frac{9c^2}{4(n+s)^2},\qquad n=0,1,2,\dots,
\end{equation}
where the parameter $s$ is given by the largest solution of the indicial equation,
\begin{equation}
\label{ind_eq}
s(s-1)=\frac{15}{4}+m_5^2R^2.
\end{equation}
Applying the relation~\eqref{s_delta}, we finally arrive at the mass spectrum of the model,
\begin{equation}
\label{s_mass}
  M_n^2=c^2-\frac{9c^2}{4(n+\Delta-3/2)^2},\qquad n=0,1,2,\dots.
\end{equation}
The spectrum~\eqref{s_mass} has an infinite number of states with the accumulation point\footnote{One can construct
holographic models with a finite upper boundary on the mass and a finite number of massive states below this boundary, such models will have the linear
background like~\eqref{SWbac2} asymptotically at large enough $z$, this possibility is discussed in~\cite{Afonin:2009xi}.} $M_\infty=c$.

Note that the positive sign of the mass parameter, $c>0$, in the action~\eqref{act} not only provides the
confining geometry~\cite{universe} but also ensures the existence of discrete spectrum by virtue of the negative
sign of Coulomb term in the potential~\eqref{eom_pot}. This situation is qualitatively different from
the standard SW model with quadratic dilaton background where the existence of discrete mass spectrum
does not depend on the sign of the background function $f(z)$ in the action of SW models~\eqref{sw}.
Another peculiar feature of the spectrum~\eqref{s_mass} is that this spectrum is gapless: The lowest canonical mass dimension of
gauge-invariant scalar operators in QCD is $\Delta=3$ (e.g., the pseudoscalar quark bilinear $\bar{q}\gamma_5q$),
the ground state in~\eqref{s_mass} is then massless,
\begin{equation}
\Delta=3:\qquad M_0=0.
\end{equation}
We recall that in the standard quadratic SW model with $f(z)=-cz^2$, the lowest mass in the scalar spectrum is (see, e.g.,~\cite{afonin2020})
\begin{equation}
\text{Scalar quadratic SW:}\qquad M_0^2=4|c|\left(\Delta-2+\frac{c}{2|c|}\right),
\end{equation}
which has a gap for both signs of $c$.

One can construct an alternative formulation of the model under consideration.
In the spirit of ``No-wall'' holographic model of Ref.~\cite{No-wall} (see also a more detailed discussion in~\cite{Afonin:2021cwo}),
let us eliminate the exponential background in the action~\eqref{act} by the field transformation
\begin{equation}
\label{transf}
  \Phi=e^{-cz}\phi.
\end{equation}
In terms of the new field $\phi$, the action~\eqref{act} takes the form
\begin{equation}
\label{act2}
  S=\int d^5x \sqrt{g}\left(g^{MN}\partial_M\phi\partial_N\phi+
  \frac{2cz^2}{R^2}\phi\pz\phi-\lb\frac{c^2z^2}{R^2}+m_5^2\rb\phi^2\right).
\end{equation}
Since the field $\phi$ should be vanishing at spatial infinity,
the second term in~\eqref{act2} can be easily integrated by parts with the result
\begin{equation}
 \frac{2c}{R^2}\int d^5x \sqrt{g}\,z^2\phi\pz\phi=\frac{3c}{R^2}\int d^5x \sqrt{g}\,z\phi^2,
\end{equation}
where the expression~\eqref{det} was used. This yields the final form of the action in terms of $\phi$,
\begin{equation}
\label{act3}
  S=\int d^5x \sqrt{g}\left(g^{MN}\partial_M\phi\partial_N\phi-m_{5,\text{eff}}^2\phi^2\right),
\end{equation}
where the effective $z$-dependent mass is given by
\begin{equation}
\label{meff}
  m_{5,\text{eff}}^2R^2 = c^2z^2-3cz+m_5^2R^2.
\end{equation}
Repeating the same steps which lead to the equation of motion~\eqref{eom}
we arrive at this equation with the effective potential
\begin{equation}
\label{eom_pot2}
  V(z)=\frac{15/4+m_{5,\text{eff}}^2R^2}{z^2}.
\end{equation}
The substitution of~\eqref{meff} into~\eqref{eom_pot2} gives the potential~\eqref{eom_pot} as expected.
Note that a similar transformation of the standard quadratic SW model will result in a \(\mathcal{O}(z^4)\) contribution
to the effective mass~\eqref{meff} (this contribution determines the slope of Regge spectrum) while
the \(\mathcal{O}(z)\) contribution will be absent~\cite{Afonin:2021cwo}.

If the 5D field $\Phi$ is massless, $m_5=0$, another alternative formulation of the model
becomes possible: The exponential background in the action~\eqref{sw} can be replaced by
a certain modification of the AdS$_5$ metric~\cite{andreev,Afonin:2021cwo},
\begin{equation}
\label{sw2}
\int d^5x \sqrt{g} \, e^{f}\mathcal{L} \quad\longrightarrow \quad \int d^5x \sqrt{\tilde{g}}\,\mathcal{L},
\end{equation}
\begin{equation}
\label{metric2}
  \tilde{g}_{MN}=e^{\frac{2f}{3}}\frac{R^2}{z^2}\,\eta_{MN}.
\end{equation}
The resulting equation of motion will be the same since in the action~\eqref{act} we will have
\begin{equation}
\sqrt{\tilde{g}}\,\tilde{g}^{MN} = \sqrt{\frac{R^{10}}{z^{10}}\,e^{\frac{10f}{3}}}\frac{z^2}{R^2}\,e^{-\frac{2f}{3}}\eta^{MN}=\sqrt{g}\,e^{f}g^{MN}.
\end{equation}
In string theory, the reformulation~\eqref{sw2} means the transition from the string frame
to the Einstein frame. The latter formulation is used for the analysis of confinement properties~\cite{Andreev:2006ct}
(the recent discussions and references are given in~\cite{Afonin:2022aqt}).

\section{Extension to the vector case}

The vector analogue of the action~\eqref{act} has the form (a general normalization factor is omitted),
\begin{equation}
\label{actv}
  S=-\int d^5x \sqrt{g}\, e^{2cz}\left(g^{MR}g^{NS}\partial_MV_N\partial_RV_S-m_5^2\,g^{NS}V_NV_S\right),
\end{equation}
with the additional condition $\partial^M V_M=0$. According to the AdS/CFT dictionary~\cite{witten,gub}, the 5D mass $m_5$ of free
vector field $V_N$ in AdS$_5$ space is determined by the canonical dimension $\Delta$ of corresponding vector
operator in dual 4D field theory via the relation
\begin{equation}
\label{s_delta2}
m_5^2R^2=(\Delta-1)(\Delta-3).
\end{equation}
The form~\eqref{actv} is convenient for generalization to fields of arbitrary integer spin $J$ (see the next Section,
the factor $(-1)^J$ then appears in front of the action). In practice, the kinetic term for vector fields is usually
written in terms of field strength $F_{MN}=\partial_MV_N-\partial_NV_M$,
\begin{equation}
\label{actv2}
  S=\int d^5x \sqrt{g}\, e^{2cz}\left(-\frac12F^{MN}F_{MN}+m_5^2V^NV_N\right).
\end{equation}
The condition $\partial^M V_M=0$ is then satisfied automatically.
The particle-like excitations along the usual $3+1$ physical coordinates are selected by the condition
\begin{equation}
\label{ax}
V_z=0.
\end{equation}
The ensuing equation of motion is
\begin{equation}
\label{eomv}
  \partial_R\lb e^{2cz}\sqrt{g}\,g^{MR}g^{NS}\partial_MV_N\rb+e^{2cz}\sqrt{g}\,m_5^2g^{NS}V_N=0.
\end{equation}
Due to the condition~\eqref{ax} (which coincides with the ``axial gauge'' in the massless case),
the polarization vector \(\veps_\nu\) has non-zero components only along the physical 4D space.
The plain-wave ansatz for particle-like excitations,
\begin{equation}
\label{pe2}
  V_\nu(x,z)=e^{iqx}v(z)\veps_\nu,\,\qquad q^2=M^2,
\end{equation}
transforms the equation~\eqref{eomv}  into
\begin{equation}
\label{eomv2}
  -q^2v-ze^{-2cz}\pz\lb\frac{e^{2cz}}{z}\pz v\rb+\frac{R^2}{z^2}m_5^2v=0.
\end{equation}
After the substitution
\begin{equation}
\label{subV}
  v=z^{1/2}e^{-cz}\psi,
\end{equation}
we obtain the Schr\"{o}dinger equation~\eqref{eom} with the potential
\begin{equation}
  V(z)=\frac{3/4+m_5^2R^2}{z^2}-\frac{c}{z}+c^2.
\end{equation}

As in the scalar case, we can get the mass spectrum using the formulas from the Appendix,
\begin{equation}
 M_n^2=c^2-\frac{c^2}{4(n+s)^2},\qquad \qquad n=0,1,2,\dots,
\end{equation}
where the parameter $s$ is the largest solution of the corresponding indicial equation,
\begin{equation}
\label{ind_eq2}
s(s-1)=\frac{3}{4}+m_5^2R^2.
\end{equation}
Making use of the holographic relation~\eqref{s_delta2}, we get the vector spectrum of the model,
\begin{equation}
\label{s_mass2}
  M_n^2=c^2-\frac{c^2}{4(n+\Delta-3/2)^2},\qquad n=0,1,2,\dots.
\end{equation}
The spectrum~\eqref{s_mass2} possesses the same accumulation point as in the scalar case, $M_\infty=c$,
but now the spectrum is gapped: Since the lowest canonical mass dimension of
gauge-invariant vector operators in QCD is $\Delta=3$ (e.g., the vector current $\bar{q}\gamma_\mu q$),
the ground state mass in~\eqref{s_mass2} is $M_0=(\sqrt{8}/3)c$.

The main outcome of AdS/CFT correspondence is a holographic recipe for calculating the correlation functions
of strongly coupled gauge theory
from a dual gravitational theory~\cite{witten,gub}. The mass spectrum must follow from the poles of corresponding
two-point correlation function. This ``direct'' way of calculating mass spectrum in the holographic approach
technically looks relatively cumbersome but the calculated correlation functions give, as a by-product,
a lot of additional physical information, for instance, one can expand the correlators at large Euclidean
momentum and compare the expansion with the Operator Product Expansion (OPE) in QCD.
To give an example within the holographic model under consideration, we consider the calculation of
two-point correlator for the vector case. For simplicity, we will assume that the vector field is massless.
According to~\eqref{s_delta2}, such a 5D field is dual to the operator of vector current with dimension  $\Delta=3$,
i.e., to the operator $J_\mu=\bar{q}\gamma_\mu q$ in the simplest case.

We will carry out the calculations in the Euclidean momentum space, \(Q^2\equiv-q^2\).
The equation of motion~\eqref{eomv2} for massless vector field takes the form
\begin{equation}
\label{eomv3}
  Q^2V-ze^{-2cz}\pz\lb\frac{e^{2cz}}{z}\pz V\rb=0.
\end{equation}
The solution \(V(Q,z)\) of this equation with the boundary condition \(V(Q,0)=1\)
gives the holographic ``bulk-to-boundary'' propagator (the solution \(v(z)\) of~\eqref{eomv2}
describing particles satisfies the boundary condition \(v(0)=0\)).
The explicit form for such a solution of Eq.~\eqref{eomv3} is
\begin{equation}
\label{sol}
  V(Q,z)=\lb c^2-\hq^2\rb\Gamma\lb-\frac{1}{2}-\frac{c}{2\hq}\rb e^{-\lb c+\hq\rb z}z^2U\lb\frac{3}{2}-\frac{c}{2\hq},3,2\hq z\rb,
\end{equation}
where $U(a,b,x)$ is the Tricomi confluent hypergeometric function and we introduced the notation
\begin{equation}
\label{not}
\hq\equiv\sqrt{c^2+Q^2}.
\end{equation}

The bulk-to-boundary propagator plays the central role:
One can show~\cite{son1,pom} that the two-point vector correlator is given by
\begin{equation}
\label{corr}
  \Pi_V(Q^2)\sim\left.-\frac{\pz V(Q,z)}{Q^2 z}\right|_{z\to0},
\end{equation}
where a constant factor is determined by the omitted normalization factor in the action~\eqref{actv}.
Thus, we just need to find the expansion of~\eqref{sol} at \(z\to0\).
The Tricomi function has the following general series representation,
\begin{multline}
\label{tric_exp}
  U(a,b,x)\underset{x\to0}{=}\frac{(-1)^b}{\Gamma(a-b+1)}\lsb
  \frac{\ln x}{(b-1)!}\sum_{k=0}^\infty\frac{(a)_kx^k}{(b)_kk!}+\right.\\\left.+
  \sum_{k=0}^\infty\frac{(a)_k\lb\psi(a+k)-\psi(k+1)-\psi(k+b)\rb x^k}{(k+b-1)!k!}-\right.\\\left.-
  \sum_{k=1}^{b-1}\frac{(k-1)!x^{-k}}{(1-a)_k(b-k-1)!}\rsb,\quad b\in\mathbb{N},
\end{multline}
where \((a)_k\) denotes the Pochhammer symbol and $\psi(x)$ is the digamma function. In our case, this representation gives
\begin{multline}
\label{exp2}
  \lb c^2-\hq^2\rb\Gamma\lb-\frac{1}{2}-\frac{c}{2\hq}\rb z^2U\lb\frac{3}{2}-\frac{c}{2\hq},3,2\hq z\rb\underset{z\to0}{=}
  1-\frac{\ln(2\hq z)}{2}\lb c^2-\hq^2\rb z^2+\\+\lb c+\hq\rb z-
  \lb c^2-\hq^2\rb\lb\frac{1}{2}\psi\lb\frac{3}{2}-\frac{c}{2\hq}\rb+\gamma-\frac{3}{4}\rb z^2+
  \mathcal{O}(z^3).
\end{multline}
Multiplying~\eqref{exp2} by the expansion of the exponent,
\begin{equation}
\label{exp1}
  e^{-\lb c+\hq\rb z}\underset{z\to0}{=}
  1-\lb c+\hq\rb z+\lb c+\hq\rb^2\frac{z^2}{2}+\mathcal{O}(z^3),
\end{equation}
we get the expansion for $V(Q,z)$,
\begin{multline}
  V(Q,z)\underset{z\to0}{=}
  1-\frac{1}{2}\lb c^2-\hq^2\rb z^2\ln(2\hq z)-\\-
  \lsb\lb c^2-\hq^2\rb\lb\frac{1}{2}\psi\lb\frac{3}{2}-\frac{c}{2\hq}\rb+\gamma-\frac{3}{4}\rb+\frac{1}{2}\lb c+\hq\rb^2\rsb z^2+
  \mathcal{O}(z^3).
\end{multline}
The corresponding series expansion of its \(z\)-derivative,
\begin{equation}
  \pz V(Q,z)\underset{z\to0}{=}
  -\lsb\lb c^2-\hq^2\rb\lb\psi\lb\frac{3}{2}-\frac{c}{2\hq}\rb+2\gamma-\frac{3}{2}\rb+\lb c+\hq\rb^2\rsb z+
  \mathcal{O}(z^2),
\end{equation}
after substitution to~\eqref{corr}, yields 
the following expression
for the vector correlator in the model,
\begin{equation}
\label{vc0}
  \Pi_V(Q^2)\sim\frac{1}{Q^2}\lsb\lb c^2-\hq^2\rb\lb\psi\lb\frac{3}{2}-\frac{c}{2\hq}\rb+2\gamma-\frac{3}{2}\rb+\lb c+\hq\rb^2\rsb,
\end{equation}
or, in terms of usual Euclidean momentum \(Q\) in the notation~\eqref{not},
\begin{equation}
\label{vc}
    \Pi_V(Q^2)\sim\frac{3}{2}-2\gamma+\lb\frac{c}{Q}+\sqrt{\frac{c^2}{Q^2}+1}\rb^2-\psi\lb\frac{3}{2}-\frac{c}{2\sqrt{c^2+Q^2}}\rb.
\end{equation}

An important remark needs to be made here. Strictly speaking, following the AdS/CFT prescriptions for holographic calculation of
correlators~\cite{witten,gub} one obtains not $\Pi_V(Q^2)$ in~\eqref{vc0} but the two-point correlation function
\begin{equation}
\label{vc0b}
\langle J_\mu J_\nu\rangle = \left(\eta_{\mu\nu}-\frac{q_\mu q_\nu}{q^2}\right)q^2\Pi_V(q^2).
\end{equation}
Note that for arbitrary operators, the term ``correlator'' refers to the scalar function which appear after extracting
the Lorentz structure, i.e., to $q^2\Pi_V(q^2)$ in the case at hand. Only in the vector case, due to the conservation of vector current,
it is customary to extract the extra factor $q^2$ in the definition~\eqref{vc0b}.
The expression~\eqref{vc0b} has an ultraviolet divergence that can be handled by the holographic renormalization~\cite{skenderis}, this
procedure is shown in detail for the case of SW model in Ref.~\cite{AS}. The r.h.s. of~\eqref{vc0b} contains $q^2$-dependent contact
terms required for regularization of infinities. After subtracting all these terms and irrelevant constants one obtains
the physical $\langle J_\mu J_\nu\rangle^\text{(subt)}$ that defines the physical two-point correlator $\Pi_V^\text{(subt)}(q^2)$,
where ``subt'' means ``subtracted''. It is easy to see that only the last term in~\eqref{vc} will be left after performing
all subtractions of unphysical contributions. Thus, the subtracted vector correlator is
\begin{equation}
\label{vcsub}
    \Pi_V^\text{(subt)}(Q^2)\sim-\psi\lb\frac{3}{2}-\frac{c}{2\sqrt{c^2+Q^2}}\rb.
\end{equation}
It is interesting to observe that the vector correlator~\eqref{vcsub} is expressed via the same digamma function $\psi$ as
the vector correlator in the standard SW model with quadratic background but with different argument:
in the case of quadratic background~\eqref{SWbac}, the correlator is~\cite{son3}
\begin{equation}
\text{Quadratic SW:}\qquad\qquad    \Pi_V^\text{(subt)}\sim-\psi\left(1+\frac{Q^2}{4\lambda^2}\right).
\end{equation}
The poles in~\eqref{vcsub} arise at negative integer values of the argument of digamma function, $\psi(-n)$, $n=0,1,2\dots$,
which appear at the following values of Euclidean momentum,
\begin{equation}
\label{s_massE}
  Q_n^2=-\left(c^2-\frac{c^2}{4(n+3/2)^2}\right).
\end{equation}
As expected, the positions of poles~\eqref{s_massE} reproduce the mass spectrum~\eqref{s_mass2} in the Minkowski space,
$M_n^2=-Q_n^2$, for $\Delta=3$.


The structure of spectrum becomes manifest if one makes the pole decomposition of two-point correlator,
\begin{equation}
\label{sum}
\Pi_V(Q^2)\sim-\sum_{n=0}^\infty\frac{F_n^2}{Q^2+M_n^2},
\end{equation}
where the decay constants for a vector meson $V_n$ with polarization $\varepsilon_\mu$ are defined by matrix
elements of vector current between vacuum and a given vector state as
\begin{equation}
\label{defF}
\langle0|J_\mu^a|V_n^b\rangle=F_n 
\delta^{ab}\varepsilon_\mu.
\end{equation}
Using the pole representation of digamma function,
\begin{equation}
\psi(x)=\text{Const}-\sum_{n=0}^\infty\frac{1}{n+x},
\end{equation}
we obtain from~\eqref{vcsub} the following pole decomposition,
\begin{equation}
\label{sum2}
\Pi_V(Q^2)\sim\sum_{n=0}^\infty\frac{\frac{c^2+Q^2}{n+3/2}+\frac{c\sqrt{c^2+Q^2}}{2(n+3/2)^2}}{Q^2+c^2-\frac{c^2}{4(n+3/2)^2}}.
\end{equation}
Here the positions of poles~\eqref{s_massE} are manifest. The numerators $F_n^2$, however, become momentum-dependent, $F_n^2(Q^2)$.
Then the decay constant $F_n$ of $n$-th state should be defined from the corresponding residue $F_n^2(Q^2_n)$ at the $n$-th pole:
$F_n=F_n(Q^2_n)$. This yields
\begin{equation}
\label{Fn}
F_n\sim\frac{c}{(n+3/2)^{3/2}},
\end{equation}
where the general proportionality coefficient remains undetermined.

The relation~\eqref{Fn} predicts the decreasing of electromagnetic decay constants with $n$,
\begin{equation}
\label{Fn2}
\frac{F_n}{F_0}=\left(\frac{3}{2n+3}\right)^{3/2}.
\end{equation}
This prediction is in  qualitative agreement with the experimental data on vector charmonia and bottomonia,
see Ref.~\cite{Afonin:2018ejx} in which the corresponding phenomenological analysis was performed for electromagnetic couplings to
$e^+e^-$ annihilation where the experimental extraction of these couplings for radially excited states is the most reliable.
In this case, the decay constants defined as~\eqref{defF} are related to the corresponding decay width as
\begin{equation}
\Gamma_{V_n\rightarrow e^+e^-}=\frac{4\pi\alpha^2F_n^2}{3M_n}.
\end{equation}
Numerically, the relation~\eqref{Fn} predicts:
$F_1/F_0\approx0.46$, $F_2/F_0\approx0.28$, $F_3/F_0\approx0.19$. The experimental data
for these ratios are~\cite{Afonin:2018ejx}: $F_1/F_0\approx0.71$, $F_2/F_0\approx0.45$, $F_3/F_0\approx0.34$.
Thus the relation~\eqref{Fn} underestimates the experimental ratios by an approximately
universal factor of 1.5.
More generically, the relation~\eqref{Fn} predicts the following rate of decoupling of $S$-wave radial states
from the $e^+e^-$ annihilation,
\begin{equation}
F_n^2=\frac{F_0^2}{(n+3/2)^3},
\end{equation}
while according to the fits in Ref.~\cite{Afonin:2018ejx} the corresponding decoupling rate can be interpolated by
\begin{equation}
\label{Fn3}
F_n^2=\frac{F_0^2}{\beta^n},
\end{equation}
where $\beta\approx0.5-0.6$ for the radial excitations of $J/\psi(1S)$ and $\Upsilon(1S)$ mesons.
We recall that the standard SW model with quadratic dilaton predicts
an $n$-independent behavior of decay constants~\cite{son2}, $F_n=F_0$, if the definition~\eqref{defF} is used.
For this reason, the description of decay constants
of heavy quarkonia within this model requires some special tunings~\cite{Grigoryan:2010pj}.
It should be also mentioned that the phenomenological law~\eqref{Fn3} may result from a nontrivial form-factor
of excited states~\cite{Afonin:2018ejx} which is not taken into account in simple models.

Finally, the correlator~\eqref{vcsub} can be expanded at large \(Q^2\),
\begin{equation}
\label{op}
\Pi_V^\text{(subt)}(Q^2)\underset{Q^2\to\infty}{\sim}\text{Const}+
   \frac{(\pi^2-8)c}{4Q}+\frac{(7\zeta(3)-8)c^2}{4Q^2}+\mathcal{O}\lb\frac{1}{Q^3}\rb.
\end{equation}
This expansion has a different structure than the OPE in QCD. First of all, there is no logarithmic
contribution $\mathcal{O}\lb\ln{Q^2}\rb$ that emerges in the leading order of perturbation theory (the parton model logarithm).
As a result, one cannot perform the ensuing matching and extract the proportionality coefficient in~\eqref{Fn}.
Also the expansion~\eqref{op}
contains odd powers of $1/Q$ which are absent in the standard OPE of correlators of QCD currents~\cite{svz}.
This is a manifestation of a specific property known from general considerations of quark-hadron duality in the large-$N_c$ limit of QCD:
An infinite radial spectrum of narrow states can reproduce the correct structure of OPE corrections to the perturbative logarithm
(i.e., an expansion in powers of $1/Q^2$) only if the spectrum is a Regge-like with possible non-linear corrections decreasing with $n$
exponentially or faster~\cite{lin2,lin3}.

\section{Extension to arbitrary integer spin}

The case of vector mesons can be formally extended to the case of tensor mesons.
The fields of arbitrary integer spin $J$ are described by the symmetric traceless tensors $\Phi_{M_1\dots M_J}$ with zero
divergence, $\partial^M\Phi_{MM_2\dots M_J}=0$.
In general, the description of higher-spin fields in the AdS space is a complex problem. The treatment of tensor fields in the
bottom-up holographic QCD requires a simplified framework to extend the computations to the cases of asymptotic AdS spaces.
The given issue is discussed in detail in Refs.~\cite{br2,br3}. In short,
several recipes have been proposed in the literature. We will follow the most used method.
The Lagrangian of free massive tensor fields in the bottom-up holographic models takes a simple form
generalizing the action~\eqref{actv} for massive vector fields,
\begin{equation}
\label{LJ}
  \mathcal{L}_{J}=g^{MN}g^{M_1N_1}\dots g^{M_JN_J}
  \partial_M\Phi_{M_1\dots M_J}\partial_N\Phi_{N_1\dots N_J}-
  m_5^2\,g^{M_1N_1}\dots g^{M_JN_J}\Phi_{M_1\dots M_J}\Phi_{N_1\dots N_J},
\end{equation}
with normal derivatives (the covariant derivatives can be replaced by the normal ones via a certain rescaling of fields~\cite{br2,br3})
and where the tensor generalization of condition~\eqref{ax} is imposed,
\begin{equation}
\label{ax2}
\Phi_{zM_2\dots M_J}=0,
\end{equation}
to select the particle-like excitations along the usual $3+1$ physical coordinates.
This condition becomes identical to the axial gauge when the higher-spin fields are described as massless gauge fields~\cite{son2}.
The condition~\eqref{ax2} eliminates additional terms in~\eqref{LJ} arising from various ways of contraction of Lorentz indices.
The 5D mass is given by the corresponding holographic relation~\cite{br3,sundrum},
\begin{equation}
\label{s_deltaJ}
m_5^2R^2=(\Delta-J)(\Delta+J-4),
\end{equation}
which generalizes the relations~\eqref{s_delta} and~\eqref{s_delta2}. We refer to Refs.~\cite{br2,br3} for the relevant details.
One can directly show (see, e.g.,~\cite{Afonin:2021cwo,afonin2010}) that in the case of usual SW model with positive quadratic dilaton,
this recipe leads to the spectrum~\eqref{spSW} which was originally obtained in~\cite{son2} using a different recipe
(namely, starting from the massless higher-spin fields in AdS$_5$ and considering the SW model with negative quadratic dilaton).

Within our model, the Lagrangian~\eqref{LJ} stays in the action
\begin{equation}
\label{act4}
  S=\int d^5x\,e^{2cz}\sqrt{g}\,\mathcal{L}_{J},
\end{equation}
where the metric is given by~\eqref{metric} and a spin-dependent general normalization factor is omitted.
The equation of motion is
\begin{equation}
  \partial_N\lb e^{2cz}\sqrt{g}\,g^{MN}g^{M_1N_1}\dots g^{M_JN_J}
  \partial_M\Phi_{M_1\dots M_J}\rb+
  e^{2cz}\sqrt{g}\,m_5^2\,g^{M_1N_1}\dots g^{M_JN_J}\Phi_{M_1\dots M_J}=0.
\end{equation}
Exploiting the standard plain wave ansatz,
\begin{equation}
  \Phi_{\mu_1\dots\mu_J}=e^{iq^\mu x_\mu}\phi(z)\veps_{\mu_1\dots\mu_J},\qquad q^\mu q_\mu=M^2,
\end{equation}
where \(\veps_{\mu_1\dots\mu_J}\) is a polarization tensor with indices along the physical 4D coordinates only,
the equation of motion transforms into
\begin{equation}
  -q^2\phi-z^{3-2J}e^{-2cz}\pz\lb\frac{e^{2cz}}{z^{3-2J}}\pz\phi\rb+
  \frac{R^2}{z^2}m_5^2\phi=0.
\end{equation}
Performing the substitution \(\phi=z^{3/2-J}e^{-cz}\psi\) we obtain
\begin{equation}
  -q^2\psi-\pzz\psi+
  \lb\frac{(J-2)^2-1/4+m_5^2R^2}{z^2}+\frac{(2J-3)c}{z}+c^2\rb\psi=0,
\end{equation}
which is the Schr\"{o}dinger equation with the spin-dependent potential
\begin{equation}
\label{VJ}
  V(z)=\frac{(J-2)^2-1/4+m_5^2R^2}{z^2}+\frac{(2J-3)c}{z}+c^2.
\end{equation}
As in the previous cases, we can immediately write the discrete spectrum,
\begin{equation}
  M_{n,J}^2=c^2-\frac{(2J-3)^2c^2}{4(n+s)^2},\qquad \qquad n=0,1,2,\dots,
\end{equation}
where $s$ is the largest solution of the equation,
\begin{equation}
\label{ind_eq3}
s(s-1)=(J-2)^2-1/4+m_5^2R^2.
\end{equation}
Using the holographic relation~\eqref{s_deltaJ}, we finally obtain the spectrum
\begin{equation}
\label{s_massJ}
  M_{n,J}^2=c^2-\frac{(2J-3)^2c^2}{4(n+\Delta-3/2)^2},\qquad n=0,1,2,\dots.
\end{equation}
The scalar spectrum~\eqref{s_mass} and the vector one~\eqref{s_mass2} are special cases of the
general formula~\eqref{s_massJ}.

When the meson states are interpolated by operators of the lowest twist, one has $\Delta=J+2$
for all mesons carrying a non-zero spin, $J>0$. Then the spectrum~\eqref{s_massJ} takes the form
\begin{equation}
\label{s_massJ2}
\Delta=J+2,\quad J>0:\qquad\qquad  M_{n,J}^2=c^2-\frac{(2J-3)^2c^2}{4(n+J+1/2)^2},\qquad n=0,1,2,\dots.
\end{equation}
The denominator here contains the same $(n+J)$-degeneracy as the string-like spectrum~\eqref{spSW}
(and as the Coulomb spectrum if $J$ is associated with the orbital angular momentum) but
this degeneracy is lifted by the $J$-dependent numerator. The same situation holds for higher twists,
$\Delta=J+k$, $k>2$. In any case, the spectrum has an obvious pathology: $M_{n,J}\rightarrow0$ at
$J\rightarrow\infty$.

The encountered non-physical behavior of the spectrum deserves a serious discussion.
It is clear from the derivation above that the problem emerges because we tacitly identified the total
angular momentum of a hadron with the fundamental spin which is not related with the space-time rotations
in the classical limit. This identification is justified for fundamental
strings. In the case of composite systems like real hadrons, the situation is more complex: The total angular momentum $\vec{J}$
represents the sum of total quark spin $\vec{s}$ and the intrinsic orbital angular momentum $\vec{l}$. In the non-relativistic
limit (which we will consider in the next Section), they are separately conserved. Only $\vec{s}$ can be handled
as a genuine spin while $\vec{l}$ describes the internal space-time orbital excitations. The latter is known to define
the space parity of a quark-antiquark pair, $P=(-1)^{l+1}$, where $l$ is the orbital quantum number.

The quark-antiquark states with intrinsic orbital angular momentum $\vec{l}$ are created in QCD
by symmetrized (and traceless) products of covariant derivatives $D_{l_i}$~\cite{br1,forkel}. The corresponding QCD operators
have the canonical dimension $\Delta=3+l$ and take the form
\begin{equation}
\mathcal{O}_{3+l}=\bar{q}\Gamma D_{\{l_1}D_{l_2}\dots D_{l_m\}} q,
\end{equation}
where 
$l=\sum_{i=1}^m l_i$ and $\Gamma$ denotes some gamma-matrix structure, e.g., $\Gamma=1$ for scalars and $\gamma_\mu$
for vectors. In the last two examples, the maximum total angular momentum will be $J=l$ and $J=l+1$, respectively.
The crux of the matter is that the operation of taking derivatives does not give new fundamental spin degrees of freedom,
hence, although the operator $\mathcal{O}_{3+l}$ formally has $l$ additional Lorentz indices, it cannot be strictly dual
to some higher-spin field in the 5D bulk. One can of course build higher-spin QCD operators describing hadron states carrying only the
fundamental spin degrees of freedom but they will correspond either to multiquark states with $l=0$ or to mixing with
glueballs. For the spin-2 case, the relevant examples are $\bar{q}\gamma_\mu q\bar{q}\gamma_\nu q$ and
$\bar{q}G_\mu^\rho G_{\rho\nu}q$, correspondingly. But such states should disappear in the large-$N_c$ limit of QCD~\cite{hoof,wit},
while the holographic approach to QCD, based on the gauge/gravity duality, has theoretical motivations in this limit only.

These arguments show that within the bottom-up holographic QCD, there are reasons to describe the spectrum of higher-spin hadron
resonances by simply changing the canonical dimension $\Delta$ in the corresponding formulas for scalars and vectors. Actually this way
was pursued by some authors either partially~\cite{br3} or completely~\cite{forkel}. For instance, in the model of Ref.~\cite{forkel}, instead of the
string-like spectrum~\eqref{spSW} the authors obtained the following phenomenological spectrum
\begin{equation}
\label{specl}
  M^2_{n,l}=4\lambda^2(n+l+1/2),\qquad n,l=0,1,2,\dots,
\end{equation}
which virtually coincides with the phenomenological spectrum of light non-strange mesons found independently in~\cite{phen4,phen5}
(see also~\cite{klempt,phenSV}) from fitting the available experimental data. In the case under consideration, the orbital excitations
of scalar and vector quark-antiquark states can be then obtained by substituting $\Delta=3+l$ into~\eqref{s_mass}
and~\eqref{s_mass2}, respectively. In the latter case, for example, the spectrum takes the following form,
\begin{equation}
\label{s_massl}
  M_{n,l}^2=c^2-\frac{c^2}{4(n+l+3/2)^2},\qquad n,l=0,1,2,\dots.
\end{equation}
We see thus that the spectrum becomes much closer to the Hydrogen-like spectrum, especially after taking the non-relativistic
limit (see the next Section). As an additional bonus, the troubles with affine connections in curved space-times and related
ambiguity with the description of higher-spin fields in modified AdS$_5$ spaces disappear since the corresponding complications
do not arise in the Lagrangians for free scalar~\eqref{act} and vector~\eqref{actv2} fields.

The remark above is not essential for the standard SW model with quadratic exponential background because it leads to a simple
replacement of $J$ in the string-like spectrum~\eqref{spSW} by $l$ plus a constant~\cite{forkel} or by a linear combination of $l$ and $J$~\cite{br3}
(more precisely, the role of $l$ in~\cite{br3} is played by the light-front internal orbital angular momentum $L$).
This can have only some phenomenological consequences when fitting experimental data. In the considered model, the problem of
holographic description of higher-spin mesons becomes more serious.

In the literature, there are many discussions on the problem of gauge-invariant separation between the
total quark spin $\vec{s}$ and the intrinsic orbital angular momentum $\vec{l}$ starting with the seminal paper~\cite{ji}
(the recent lattice QCD study of the spin problem is reviewed in~\cite{Lin}). Unfortunately, they hardly can help because, first,
these discussions refer to essentially different class of hadrons --- the stable nucleons for which a relevant experimental check
is possible, second, even for the nucleons the final theoretical picture for the spin structure is far from established.

It should also be added that there is an alternative formulation of SW model via introduction of effective $z$-dependent mass instead of
the dilaton background or modified AdS metric~\cite{Afonin:2021cwo,No-wall}. This formulation has more phenomenological freedom.
To demonstrate this possibility let us replace the action~\eqref{act4} by an action in pure (Poincar\'{e} patch of) AdS$_5$ space,
\begin{equation}
  S=\int d^5x\,\sqrt{g}\,\mathcal{L}_{J,\text{eff}},
\end{equation}
in which the effective 5D mass has the following $z$-dependence\footnote{Note that the effect of AdS affine connections in covariant derivatives
can be reabsorbed into an effective $\mathcal{O}(z^2)$-dependence of the 5D mass term, see a detailed discussion in~\cite{br3,br2}.},
\begin{equation}
\label{meff2}
  m_{5,\text{eff}}^2R^2 = c^2z^2+kcz+m_5^2R^2.
\end{equation}
The effective potential~\eqref{VJ} will be then replaced by
\begin{equation}
\label{eom_pot3}
  V(z)=\frac{(J-2)^2-1/4+m_{5,\text{eff}}^2R^2}{z^2}.
\end{equation}
As is clear from the comparison of~\eqref{eom_pot3} with~\eqref{VJ}, the specific choice $k=2J-3$ corresponds to the model considered above.
The given case can be obtained by applying the field transformation~\eqref{transf} in the action~\eqref{act4}. This was demonstrated in
Section~2 for the scalar field, the resulting effective mass~\eqref{meff} is a particular case of~\eqref{meff2} for $k=(2J-3)|_{J=0}=-3$.
However, in building phenomenological models we can choose any $k<0$ in~\eqref{meff2}, the spectrum~\eqref{s_massJ} will have a general form
\begin{equation}
\label{s_massJ3}
  M_{n,J}^2=c^2-\frac{k^2c^2}{4(n+\Delta-3/2)^2},\qquad n=0,1,2,\dots.
\end{equation}
In particular, the simplest choice of $k=-1$ leads to the Hydrogen-like spectrum~\eqref{s_massl} (in which $\Delta=3+l$ was taken).
The aforementioned pathology in~\eqref{s_massJ2} at high spins disappears if $k$ is a constant or at least grows with $J$
not faster than the linear power of $J$. Many other models can be obtained within such a phenomenological approach.
For instance, the authors of Ref.~\cite{forkel} proposed to introduce SW-like models via the infrared $z$-dependent correction
to the canonical dimensions of QCD operators in the form $\Delta\rightarrow\Delta+\lambda^2z^2$. The given phenomenological
prescription is tantamount to a specific recipe of introducing $z$-dependence into the 5D mass~\eqref{s_deltaJ}.
The Regge spectrum~\eqref{specl} was obtained in~\cite{forkel} within the framework of this particular recipe.
In our case, an analogous prescription would be $\Delta\rightarrow\Delta+cz$. It is easy to
show that such a model would correspond to the choice $k=-2(\Delta-2)$ in~\eqref{s_massJ3}.

Within the framework of light-front holographic method~\cite{br3} --- a fruitful spin off of the AdS/QCD approach ---
the dimensions of QCD operators are enumerated by $\Delta=L+2$, where $L$ is the light-front internal orbital angular momentum.
The obtained mass formulas can be easily adjusted for this case.

In view of the discussion above, it would be useful to clarify the meaning of changing the operator dimension $\Delta$
in AdS/QCD models. We remind the reader that the bulk fields propagating in AdS$_5$ are classified by unitary, irreducible
representations of $\mathcal{P}_{\text{AdS}_5}$ --- the group of isometries of AdS$_5$ space. The corresponding representations
$D(m_5^2,J_1,J_2)$ are characterized by three numbers --- minimal energy squared $m_5^2$ and two spins $J_1$ and $J_2$.
The local operators of the boundary CFT are classified by unitary, irreducible
representations of $SO(4,2)$ --- the conformal group of 4D flat space. The first number characterizing its representations
$D(\Delta,J_1,J_2)$ is the conformal dimension of an operator. The isomorphism of groups $\mathcal{P}_{\text{AdS}_5}$ and
$SO(4,2)$ allows to map the corresponding 5D fields to 4D local operators. The relation~\eqref{s_deltaJ} appears
as a result of this mapping\footnote{To be more precise, $m_5^2$ takes discrete values because in the AdS/CFT correspondence
the fields live in the space $\text{AdS}_5\times\text{S}^5$ and thus the 5D fields in AdS$_5$ represent Kaluza-Klein harmonics
on the five-dimensional sphere. It turns out that these harmonics can be ``enumerated'' by $\Delta$.}
for the case $J_1=J_2=J/2$ (symmetric tensors). All this is valid only for an unbroken conformal
symmetry. In a theory with confinement, when $SO(4,2)$ is broken to the Poincar\'{e} group $\mathcal{P}$ and a mass scale emerges,
one expects that the mapping is not destroyed but the relation~\eqref{s_deltaJ} should be ``distorted''. A viable
phenomenological possibility is a simple shift of $\Delta$ discussed above (in a sense, a redefinition of what we call
``$\Delta$'' in strongly coupled QCD).

Finally we note that at large angular momentum $l$ the spectrum~\eqref{s_massl} approaches its accumulation point as
\begin{equation}
  M_{n,l}-M_{n,\infty}\sim\frac{c}{l^2},\qquad l\gg1.
\end{equation}
The same qualitative behavior has been recently observed in the spectrum of open string scattering amplitude
for strings ending on a D-brane in the AdS space~\cite{Maldacena:2022ckr}.

\section{A conjecture for incorporation of heavy-quark mass into the SW holographic model}

The dilaton background in the action of SW holographic model~\eqref{sw} is defined by the function $f(z)$.
Consider the usual SW holographic model with quadratic background and unfixed sign of this background,
\begin{equation}
\label{b1}
f=\pm\lambda^2 z^2.
\end{equation}
The issue of sign in~\eqref{b1} is somewhat controversial in the literature. On the one hand, as was emphasized in~\cite{son3},
the positive sign leads to certain unacceptable features, on the other hand the positive sign is required\footnote{Also in the
case of negative sign in~\eqref{b1}, the mass spectrum does not depend on spin if the method of previous Section is applied:
one would obtain $M^2_{n,J}=4\lambda^2(n+1)$ for any non-zero spin $J$~\cite{br3,afonin2010}.} by the confinement in the sense
of holographic Wilson loop~\cite{Andreev:2006ct}. This latter property will be used in what follows.
A detailed comparison of SW models with different signs in~\eqref{b1} is contained in Refs.~\cite{Afonin:2021cwo,br3}.
The main conclusion is that the two versions of the sign appear to correspond to two different models, each with its own 
advantages and disadvantages in the phenomenological applications. In particular, the method of introducing an arbitrary spin 
used in the previous section requires a positive sign~\cite{br3}. In the discussion below, we will temporally keep the sign arbitrary.

The SW model with linear exponential background as in the action~\eqref{act} can be formally obtained from~\eqref{b1}
if we make a shift in the holographic coordinate,
\begin{equation}
\label{shift}
z\rightarrow z\pm b,\qquad b>0,
\end{equation}
which gives in~\eqref{b1}
\begin{equation}
\label{b2}
f_b=\pm\lambda^2 (z\pm b)^2=\pm\lambda^2 (z^2\pm 2bz+b^2),
\end{equation}
and then consider a formal limit where the first term in~\eqref{b2} is neglected,
\begin{equation}
\label{lim}
b\gg z/2.
\end{equation}
The last terms in~\eqref{b2} can be attributed to a change of general normalization factor in front of the action
(this factor was omitted in our definition of the SW model). The linear background $f=2cz$ considered in our work emerges
after the identification
\begin{equation}
\label{id}
c=\lambda^2b.
\end{equation}

We are going to interpret the shift~\eqref{shift} as a phenomenological way for introduction of new mass scale, $\mathcal{M}=1/b$,
that is related with the threshold of production of heavy quark-antiquark pair. Superficially, the limit~\eqref{lim} makes no sense
since the coordinate $z>0$ is unbounded. In the holographic theories with confinement, however, an effective maximum
value for $z$ appears~\cite{Andreev:2006ct}. Then the limit~\eqref{lim} happens to make physical sense,
\begin{equation}
\label{lim2}
b\gg z_\text{max}/2.
\end{equation}
The value of $z_\text{max}$ can be expressed in terms of $\lambda$ and $b$ for any specific case.

Consider the simplest case --- the SW model for massless scalar field. The first step is to rewrite a model with
dilaton background as a model with modified metric (i.e., to make transition to the Einstein frame) as is shown
in~\eqref{sw2} and~\eqref{metric2}. Then $z_\text{max}$ corresponds to the minimum of the component
$\tilde{g}_{00}$~\cite{Kinar:1998vq} (see also~\cite{Afonin:2022aqt,br3,afonin2023}), i.e., to the minimum of gravitational potential energy.
The minimum of the function (see~\eqref{metric2},~\eqref{b1} and~\eqref{shift})
\begin{equation}
\label{comp00}
  \tilde{g}_{00}=e^{\frac23\lambda^2 (z+b)^2}\frac{R^2}{z^2},
\end{equation}
is at
\begin{equation}
\label{z0}
  z_0=\frac12\left(-b+\sqrt{b^2+6/\lambda^2}\right),
\end{equation}
where the positive root was chosen because $z>0$. Identifying $z_\text{max}=z_0$ and substituting it to~\eqref{lim2},
we will have
\begin{equation}
\label{lim3}
b\gg \frac14\left(-b+\sqrt{b^2+6/\lambda^2}\right),
\end{equation}
that leads to
\begin{equation}
\label{lim4}
b\gg \frac{1}{2\lambda}.
\end{equation}
In terms of our mass parameter $c$ in~\eqref{id}, we get
\begin{equation}
\label{lim5}
c\gg \frac{\lambda}{2}.
\end{equation}
An average phenomenological value of the slope in the linear Regge spectrum~\eqref{spSW} extracted
from the known light-meson spectrum lies near $4\lambda^2\approx1.1$~GeV$^2$~\cite{bugg,phen4,phen5,klempt}.
Then phenomenologically the inequality~\eqref{lim5} reads
\begin{equation}
\label{lim6}
c\gg 260\,\text{MeV}.
\end{equation}
This estimate shows that the scalar SW model with linear dilaton can indeed arise
as a limit of the ``shifted'' scalar SW model with quadratic dilaton background if the mass parameter
of the former model satisfies, roughly speaking, $c\gg \Lambda_\text{QCD}$. It is therefore natural
to relate the mass parameter $c$ with a heavy-quark mass scale.

This point can be further substantiated by matching the Coulomb-like spectrum in our model with
with the Coulomb spectrum arising from the one-gluon exchange between heavy quark and antiquark at very short distances.
In the latter case, the non-relativistic spectrum of discrete energies has the well-known Hydrogen-like form (see the Appendix),
\begin{equation}
\label{nonrel}
E_n'=-\frac{\tilde{\alpha}_s^2M_Q}{4(n+l+1)^2},\qquad n,l=0,1,2,\dots,
\end{equation}
where $\tilde{\alpha}_s=\frac43\alpha_s$ and $M_Q$ is the quark mass. We count energies from zero, so there is no
additive constant in~\eqref{nonrel}. The total energy $E$ of heavy quarkonium $\bar{Q}Q$ contains a contribution $E_\text{con}$
from the (approximately linear) confinement potential. Let us subtract the latter contribution, denoting
$\check{E}=E-E_\text{con}$ we have
\begin{equation}
\check{E} = 2M_Q + E'.
\end{equation}
For the square of this energy in non-relativistic approximation we thus get $\check{E}^2\simeq 4M_Q^2+ 4M_QE'$.
With the help of~\eqref{nonrel} we then obtain
\begin{equation}
\label{relE}
\check{E}_n^2=4M_Q^2-\frac{\tilde{\alpha}_s^2M_Q^2}{(n+l+1)^2},\qquad n,l=0,1,2,\dots.
\end{equation}
The spectrum~\eqref{relE} is derived for the case of interactions carried out through exchange of massless vector particles. It is natural to
match~\eqref{relE} with our spectrum~\eqref{s_massl} for orbital and radial excitations of vector mesons. The matching yields
\begin{equation}
\label{matching}
c=2M_Q,\qquad\qquad \tilde{\alpha}_s=1.
\end{equation}
Hence, the mass parameter $c$ acquires the physical meaning of the threshold for production of a heavy quark-antiquark pair.
The estimate~\eqref{lim6} turns into the following condition,
\begin{equation}
\label{lim7}
M_Q\gg 130\,\text{MeV},
\end{equation}
i.e., $M_Q\gg m_s$, where $m_s\simeq130$~MeV is the mass of strange quark taken at the scale of 1~GeV in
the $\overline{\text{MS}}$ scheme~\cite{pdg}. In this regard, we get a remarkable phenomenological self-consistency.

The second equality in~\eqref{matching} gives $\alpha_s=0.75$. This value lies near the onset of strong
coupling regime where the perturbative considerations fail. The strong coupling $\alpha_s$ in QCD is known
to be running with distances. In the model potentials for heavy quarkonia like the Cornell potential, one usually
deals with an effective coupling averaged over some interval. On the other hand, there are a lot of indications on freezing
of $\alpha_s$ at low energies (see, e.g., the most recent review~\cite{Deur:2023dzc}). Various models and methods describing this non-perturbative
effect predict freezing of $\alpha_s$ near a typical value close to $\alpha_s\simeq0.7$ (see, e.g.,~\cite{Mattingly:1993ej}).
Note also that the classical QCD sum rules lead to an
observable value of the $\rho$-meson mass if one sets $\alpha_s=0.7$ as well~\cite{svz}. The numerical result for $\alpha_s$
following from our matching looks thus reasonable.

If one uses the formulation of model in terms of $z$-dependent 5D mass~\eqref{meff2} then matching with
the Hydrogen-like spectrum~\eqref{s_massJ3} gives $\tilde{\alpha}_s=|k|$ with the free parameter $k$.

Finally we note that within the standard holographic approach, the current quark mass appears as
the coefficient of the first order expansion term of the scalar field dual to the quark bilinear operator $\bar{q}q$~\cite{gub}. 
The incorporation of heavy-quark mass via the phenomenological matching~\eqref{matching} is not directly related
with this holographic prescription. A study of possible relation between these two ways is planned for the future.

\section{Towards building a new holographic model for heavy quarkonia}

The considered model was conceived as a mean for developing a new way for incorporation of heavy quark mass rather than for quantitative 
description of heavy meson spectra since the spectrum of heavy quarkonia is not Coulomb-like. For this reason, the proposed model must 
be regarded as a possible constituent part of more elaborated holographic models describing the heavy quarkonia. The construction of such models
is a separate task which is planned for the future. In the present note, we will briefly propose a particular example. 

The simplest possibility for extension of our model seems just to consider the shifted dilaton~\eqref{b2} as it is.
Considering the general case of arbitrary spin, it is easy to show that the potential~\eqref{VJ} takes then the following form (according to our 
discussions above, we choose the positive sign in~\eqref{b2}),
\begin{equation}
\label{VJ2}
  V(z)=\frac{\eta_1}{z^2}+\frac{\eta_2}{z}+\eta_3 z + \eta_4 z^2 + \eta_5,
\end{equation}
where
\begin{equation}
\label{VJ3}
  \eta_1=(J-2)^2-\frac14+m_5^2R^2,\quad\!\!\! \eta_2=\lambda^2b(2J-3),\quad\!\!\! \eta_3=2\lambda^4b,\quad\!\!\! \eta_4=\lambda^4,\quad\!\!\! \eta_5=2\lambda^2(J-1)+\lambda^4b^2.
\end{equation}
The Schr\"{o}dinger equation~\eqref{eom} with the potential~\eqref{VJ2} has no a textbook analytical solution 
but it can be analyzed analytically with the help of the Nikiforov–Uvarov method. The corresponding analysis was carried out in the recent
paper~\cite{Purohit:2022mwu}, where a potential with the structure of~\eqref{VJ2} was obtained as an approximation for the linear plus modified Yukawa 
potential proposed for description of the quark-antiquark interactions (see Eq.~(4) in Ref.~\cite{Purohit:2022mwu}). Since in the bottom-up
holographic descriptions of the quark-antiquark states the holographic coordinate $z$ has the physical interpretation of a measure of the interquark distance $r$~\cite{br3},
the analysis of Ref.~\cite{Purohit:2022mwu} is directly applicable to our case. This analysis shows that the potential~\eqref{VJ2} can give a reasonable 
quantitative description of the experimental heavy meson spectra.

\section{Conclusions}

We carried out a thorough analysis of a holographic soft-wall model with a linear dilaton.
This model can be obtained from the standard holographic SW model with quadratic dilaton
by making a shift of holographic coordinate, $z\rightarrow z+b$, and taking the limit of very
large $b$. The model has a Hydrogen-like spectrum of massive states. 
We demonstrated that the parameter $b$ should be proportional to the heavy-quark threshold.
This result provides thus a universal recipe for embedding the heavy-quark mass into the SW holographic approach.

The electromagnetic coupling constants predicted by the model are shown to be decreasing with the radial
number. This is qualitatively consistent with the experimental observations. It should be recalled that
the SW model with quadratic dilaton predicts constant couplings, which is inconsistent with the phenomenology~\cite{Afonin:2018ejx}.

The form of the linear SW model turns out to be more restricted than the form of the standard quadratic SW model:
the sign of dilaton background is fixed and the description of higher spin mesons has less phenomenological freedom.

It is curious to observe that the spectrum of the considered linear holographic SW model is somewhat similar to the
spectrum of open string scattering amplitude in the AdS space which has been found recently~\cite{Maldacena:2022ckr}.
The usual quadratic SW model possesses an analogous property: its linear Regge spectrum is similar to the
spectrum of open string scattering amplitude in the flat space. A further development of this analogy could be interesting.

The considered model may find applications as an important component in the construction of bottom-up holographic models aimed at a
quantitative description of the physics related with heavy flavors (spectroscopy and structure of quarkonia, QCD thermodynamics, etc.).
The SW model with shifted dilaton considered in the Section~6 represents a possible example of such an improved model.

\section*{Acknowledgements}

This research was funded by the Russian Science Foundation grant number 21-12-00020.

\section*{Appendix}

\renewcommand\theequation{A.\arabic{equation}}
\setcounter{equation}{0}

The Coulomb problem for two particles of equal mass $m$ and of opposite charges $e$ and $-e$ is defined
by the following Schr\"{o}dinger equation (in units of $\hbar=1$),
\begin{equation}
-y''+\left[-\frac{m\alpha}{r}+\frac{l(l+1)}{r^2}\right]y=mE\,y,
\end{equation}
where $\alpha$ is the fine structure constant and $l=0,1,2,\dots$ is the orbital quantum number.
In terms of $y(r)$, the radial wave function in the Coulomb problem is $R(r)=y(r)/r$.
The spectrum of discrete energies is given by (see, e.g.,~\cite{landau} or any other book on Quantum Mechanics)
\begin{equation}
\label{En}
E=-\frac{m\alpha^2}{4(n+s)^2}=-\frac{m\alpha^2}{4(n+l+1)^2}\equiv-\frac{m\alpha^2}{4N^2}, \qquad n=0,1,2,\dots.
\end{equation}
Here the value of parameter $s$ is determined from the following indicial equation,
\begin{equation}
s(s-1)=l(l+1),
\end{equation}
which has the solutions $s=l+1$ and $s=-l$. The physically acceptable normalizable wave functions
appear only at $s=l+1$. The corresponding eigenfunctions, normalized as $\int_0^\infty y^2dr=1$,
read as follows~\cite{landau}
\begin{equation}
y=\frac{1}{N}\sqrt{\frac{(N-l-1)!}{[(N+l)!]^3}}\,e^{-\rho/2}\rho^{l+1}L^{2l+1}_{N+l}\left(\rho\right), \qquad \rho= \frac{m\alpha}{N}\,r,    
\end{equation}
where $L^a_b$ are associated Laguerre polynomials. Their asymptotics near the origin is~\cite{landau}
\begin{equation}
\label{asym0}
y(r)|_{r\rightarrow0}\simeq r^{l+1} \frac{2^{l+1}}{N^{2+l}(2l+1)!}  \sqrt{\frac{(N+1)!}{(N-l-1)!}}.
\end{equation}

\end{document}